\documentclass[aps,axodraw,twocolumn,prl]{revtex4}
\usepackage{fancyhdr}
\usepackage{fancybox}
\usepackage{graphicx}
\usepackage{color}
\usepackage{soul}
\usepackage{latexsym}

\def\hu{h_u}
\def\hd{h_d}

\def\mh{m_h}
\def\hi{h_1}
\def\hii{h_2}
\def\ai{a_1}
\def\mhi{m_{h_1}}

\def\mai{m_{a_1}}
\def\mueff{\mu_{\rm eff}}

\def\noi{\noindent}

\def\lam{\lambda}

\def\mh{m_h}

\def\mhusq{m_{H_u}^2}
\def\mhdsq{m_{H_d}^2}
\def\mssq{m_S^2}
\def\mqsq{m_Q^2}
\def\musq{m_U^2}
\def\mdsq{m_D^2}
\def\mlsq{m_L^2}
\def\mesq{m_E^2}

\def\h{h}
\def\mh{m_{h}}

\def\hbar{\overline h}
\def\lam{\lambda}

\def\mz{m_Z}

\def\hi{h_i^0}
\def\mhi{m_{\hi}}

\def\h{h}
\def\mh{m_{\h}}

\def\lam{\lambda}

\def\tauptaum{\tau^+\tau^-}

\def\lsim{\mathrel{\raise.3ex\hbox{$<$\kern-.75em\lower1ex\hbox{$\sim$}}}}
\def\gsim{\mathrel{\raise.3ex\hbox{$>$\kern-.75em\lower1ex\hbox{$\sim$}}}}
\def\ifmath#1{\relax\ifmmode #1\else $#1$\fi}

\def\vev#1{\langle #1 \rangle}
\def\lam{\lambda}

\def\mhi{m_{h_1}}

\def\eg{{\it e.g.}}

\def\stopone{\wt t_1}
\def\stoptwo{\wt t_2}

\def\mstopone{m_{\stopone}}
\def\mstoptwo{m_{\stoptwo}}
\def\mstopmean{\sqrt{\mstopone \mstoptwo}}

\def\susy{{\rm SUSY}}

\def\eg{{\it e.g.}}

\def\mstopone{m_{\stopone}}

\def\hl{h}

\def\ha{A}

\def\mhl{m_{\hl}}

\def\mha{m_{\ha}}

\def\tanb{\tan\beta}

\def\mz{m_Z}

\def\wt{\widetilde}

\def\MPL #1 #2 #3 {{\sl Mod.~Phys.~Lett.}~{\bf#1} (#3) #2}
\def\NPB #1 #2 #3 {{\sl Nucl.~Phys.}~{\bf #1} (#3) #2}
\def\PLB #1 #2 #3 {{\sl Phys.~Lett.}~{\bf #1} (#3) #2}
\def\PR #1 #2 #3 {{\sl Phys.~Rep.}~{\bf#1} (#3) #2}
\def\PRD #1 #2 #3 {{\sl Phys.~Rev.}~{\bf #1} (#3) #2}
\def\PRL #1 #2 #3 {{\sl Phys.~Rev.~Lett.}~{\bf#1} (#3) #2}
\def\RMP #1 #2 #3 {{\sl Rev.~Mod.~Phys.}~{\bf#1} (#3) #2}
\def\ZPC #1 #2 #3 {{\sl Z.~Phys.}~{\bf #1} (#3) #2}
\def\IJMP #1 #2 #3 {{\sl Int.~J.~Mod.~Phys.}~{\bf#1} (#3) #2}
\def\NIM #1 #2 #3 {{\sl Nucl.~Inst.~and~Meth.}~{\bf#1} {#3} #2}

\def\lam{\lambda}

\def\tauptaum{\tau^+\tau^-}

\def\gam{\gamma}

\def\anti{\overline}
\def\epem{e^+e^-}

\def\eg{{\it e.g.}}

\def\anti{\overline}

\def\ai{a_1}

\def\mai{m_{\ai}}

\def\gev{~{\rm GeV}}
\def\tev{~{\rm TeV}}

\def\hi{\h_1}
\def\hii{\h_2}

\def\mhi{m_{\hi}}

\newcommand{\nc}{\newcommand}
\nc{\beq}{\begin{equation}}   \nc{\eeq}{\end{equation}}
\nc{\bea}{\begin{eqnarray}}   \nc{\eea}{\end{eqnarray}}
\nc{\baa}{\begin{array}}      \nc{\eaa}{\end{array}}
\nc{\bit}{\begin{itemize}}    \nc{\eit}{\end{itemize}}
\nc{\ben}{\begin{enumerate}}  \nc{\een}{\end{enumerate}}
\nc{\bce}{\begin{center}}     \nc{\ece}{\end{center}}
\def\beqa{\begin{eqnarray}}
\def\eeqa{\end{eqnarray}}
\def\bed{\begin{description}}
\def\eed{\end{description}}

\def\mhi{m_{h_1}}

\def\eg{{\it e.g.}}

\def\tanb{\tan\beta}

\def\simle{
    \mathrel{\rlap{\raise 0.511ex 
        \hbox{$<$}}{\lower 0.511ex \hbox{$\sim$}}}}

\def\slashchar#1{\setbox0=\hbox{$#1$}           
   \dimen0=\wd0                                 
   \setbox1=\hbox{/} \dimen1=\wd1               
   \ifdim\dimen0>\dimen1                        
      \rlap{\hbox to \dimen0{\hfil/\hfil}}      
      #1                                        
   \else                                        
      \rlap{\hbox to \dimen1{\hfil$#1$\hfil}}   
      /                                         
   \fi}

\def\lam{\lambda}

\def\ls#1{\ifmath{_{\lower1.5pt\hbox{$\scriptstyle #1$}}}}
\def\lss#1{\ifmath{^{\,\lower2.5pt\hbox{$\scriptstyle #1$}}}}

\tighten \preprint{UCD-HEP-???}
\begin{document}
\title{\Large\bf 
Escaping the Large Fine-Tuning and Little Hierarchy Problems
in the Next to Minimal Supersymmetric Model and {\boldmath $h\to aa$} Decays
\vspace*{-.1in}}

\author{
Radovan Derm\' \i \v sek
and John F. Gunion
\vspace*{-.1in}
}
\address{
Department of Physics, University of California at Davis, Davis, CA 95616} 
\begin{abstract}
We demonstrate that the NMSSM can have small fine-tuning
and modest light stop mass
while still evading all experimental constraints. For small $\tanb$
(large $\tanb$),
the relevant scenarios are such
that there is always (often) 
a SM-like Higgs boson that decays to
two lighter --- possibly much lighter --- pseudoscalar Higgses.
\end{abstract}
\maketitle
\vspace*{-.2in}
\thispagestyle{empty}

In the CP-conserving Minimal Supersymmetric Model
(MSSM), large soft-supersymmetry-breaking mass parameters
are required in order that the one-loop corrections
to the tree-level prediction for the lightest Higgs boson
($\mhl\leq \mz$) increase $\mhl$ sufficiently to avoid conflict
with lower bounds from LEP data. The large size of these 
soft-SUSY breaking masses
compared to the weak scale, the natural
scale where supersymmetry is expected, is termed 
the little-hierarchy problem. This hierarchy
implies that  a substantial amount of fine-tuning of the MSSM
soft-SUSY breaking parameters is needed.    
The severity of these problems has led to
a variety of alternative approaches. For instance,
little Higgs models \cite{Arkani-Hamed:2002qy} can be less fine tuned. Or, one can argue that
large fine-tuning is not so bad, as in ``split-supersymmetry''
\cite{Arkani-Hamed:2004fb}.  In this letter, 
we show that the Next to Minimal Supersymmetric Model (NMSSM
\cite{allg}) can avoid or at least ameliorate
the fine-tuning  and little hierarchy problems. In addition, we 
find that parameter choices that are consistent with all LEP
constraints and that yield small
fine-tuning at small $\tanb$ (large $\tanb$) are nearly always (often)
such that there is a relatively light SM-like CP-even Higgs boson that
decays into two light, perhaps very light, pseudoscalars.
Such decays 
dramatically complicate the Tevatron and LHC searches for Higgs
bosons.

The NMSSM is very attractive in its own right.
It provides a very elegant solution to the $\mu$ problem of the MSSM 
via the introduction of a singlet superfield $\widehat{S}$. 
For the
simplest possible scale invariant form of the superpotential,
the scalar component of $\widehat{S}$ naturally acquires a vacuum
expectation value of 
the order of the \susy\ breaking scale, giving rise to a value of $\mu$ 
of order the electroweak scale. 
The NMSSM is the simplest
supersymmetric extension of the standard model in which the electroweak
scale originates from the \susy\ breaking scale only. 
A possible cosmological domain wall problem \cite{abel1} can be avoided
by introducing suitable non-renormalizable operators \cite{abel2} that
do not generate dangerously large singlet tadpole diagrams
\cite{tadp}.
Hence, the phenomenology of the NMSSM deserves to be studied at least as
fully and precisely as that of the MSSM. 

Radiative corrections to the Higgs masses have been
computed \cite{radcor1,radcor2,yeg,higrad2} and basic phenomenology
of the model has been studied \cite{higsec1}.
The NMHDECAY program \cite{Ellwanger:2004xm} allows easy exploration
of Higgs phenomenology in the NMSSM. In particular, it allows for the possibility
of Higgs to Higgs pair decay modes 
(first emphasized in \cite{Gunion:1996fb}
and studied later in \cite{Dobrescu:2000jt}) and
includes the associated modifications of LEP limits.
Of greatest relevance are
$\h\to aa$ decays, where $h$ is a SM-like CP-even
Higgs boson and $a$ is a (mostly singlet) CP-odd Higgs boson.
The relevant limits come from the
analysis \cite{DELPHI} of the $Zh\to Zaa\to Zb\anti b b\anti b$
channel  and the analysis
\cite{Opal3} of the $Zh\to Zaa\to Z\tauptaum\tauptaum$ channel.  
The weaker
nature of the limits from LEP on such scenarios will play an
important
role in what follows. 

The extent to which there is a no-lose theorem for
NMSSM Higgs discovery at the LHC
has arisen as an important topic
\cite{Gunion:1996fb,Ellwanger:2001iw,Ellwanger:2003jt,Ellwanger:2004gz,higsec3}.
In particular, it has been found that the Higgs to Higgs pair
decay modes
can render inadequate the
usual MSSM Higgs search modes that give rise to a
no-lose theorem for MSSM Higgs discovery at the LHC. 
And, it is
by no means proven that the Higgs to Higgs pair modes 
are directly observable
at the LHC, although there is some hope \cite{Ellwanger:2003jt,Ellwanger:2004gz}.

Earlier
discussions of fine-tuning in the NMSSM have been given in
\cite{bast,King:1995vk}.

We very briefly review the NMSSM. Its particle content differs from
the MSSM by the addition of one CP-even and one CP-odd state in the
neutral Higgs sector (assuming CP conservation), and one additional
neutralino.  We will follow the conventions of
\cite{Ellwanger:2004xm}.  Apart from the usual quark and lepton Yukawa
couplings, the scale invariant superpotential is 
\vspace*{-.07in}
\beq \label{1.1}
\lambda \ \widehat{S} \widehat{H}_u \widehat{H}_d + \frac{\kappa}{3} \ 
\widehat{S}^3 
\vspace*{-.11in}
\eeq 
\noi depending on two dimensionless couplings
$\lambda$, $\kappa$ beyond the MSSM.  [Hatted (unhatted) capital letters denote
superfields (scalar superfield components).]  The associated trilinear soft terms are 
\vspace*{-.1in}
\beq \label{1.2}
\lambda A_{\lambda} S H_u H_d + \frac{\kappa} {3} A_\kappa S^3 \,. 
\vspace*{-.1in}
\eeq
The final two input parameters are 
\vspace*{-.1in}
\beq \label{1.3} \tan \beta = h_u/h_d\,, \quad \mu_\mathrm{eff} = \lambda
s \,, 
\vspace*{-.07in}
\eeq 
where $h_u\equiv
\vev {H_u}$, $h_d\equiv \vev{H_d}$ and $s\equiv \vev S$.
These, along with $\mz$, can be viewed as
determining the three \susy\ breaking masses squared for $H_u$, $H_d$
and $S$ (denoted $\mhusq$, $\mhdsq$ and $\mssq$) 
through the three minimization equations of the scalar potential.

Thus, as compared to the three 
independent parameters needed in the 
MSSM context (often chosen as $\mu$, $\tan \beta$ and $M_A$), the
Higgs sector of the NMSSM is described by the six parameters
\vspace*{-.1in}
\beq \label{6param}
\lambda\ , \ \kappa\ , \ A_{\lambda} \ , \ A_{\kappa}, \ \tan \beta\ ,
\ \mu_\mathrm{eff}\ .
\vspace*{-.1in}
  \eeq
We will choose sign conventions for the fields
such that $\lambda$ and $\tan\beta$ are positive, while $\kappa$,
$A_\lambda$, $A_{\kappa}$ and $\mu_{\mathrm{eff}}$ should be allowed
to have either sign. 
In addition, values must be input for the gaugino masses 
and for the soft terms related to the (third generation)
squarks and sleptons that contribute to the
radiative corrections in the Higgs sector and to the Higgs decay
widths.

Sample discussions of the fine-tuning issues for the MSSM appear in
\cite{finetuning}. We will define 
\beq F={\rm Max}_a F_a \equiv {\rm
  Max}_a\left|{d\log \mz\over d\log a}\right|\,, 
\eeq 
where the parameters $a$ comprise $\mu$, $B_\mu$ and
 the other GUT-scale soft-SUSY-breaking
parameters.  (In some papers, ${d\log \mz^2\over d\log a}$ is employed.) 
 In our approach, we choose
$\mz$-scale values for all the squark soft masses squared, the gaugino
masses, $M_{1,2,3}(\mz)$, $A_t(\mz)$ and
$A_b(\mz)$ 
(with no requirement of universality at the GUT scale). We also
choose $\mz$-scale values for $\tanb$, $\mu$ and $\mha$;
these uniquely determine $B_\mu(\mz)$.
The vevs
$h_u$ and $h_d$ at scale $\mz$ are fixed by $\tanb$ and $\mz$ via $\mz^2=\anti g^2(h_u^2+h_d^2)$ (where
$\anti g^2=g^2+g^{\prime\,2}$). Finally,  $\mhusq(\mz)$
and $\mhdsq(\mz)$ are determined by 
the two potential minimization conditions. 
We then evolve all parameters
to the MSSM GUT scale (including $\mu$ and $B_\mu$).  Next, we shift
each of the GUT-scale parameters in turn, evolve back down to scale
$\mz$, and reminimize the Higgs potential using the shifted values of
$\mu$, $B_\mu$, $\mhusq$ and $\mhdsq$. This gives new values for $h_u$
and $h_d$ yielding new values for $\mz$ and $\tanb$.

\begin{figure}[h]
\centerline{\includegraphics[width=2.4in,angle=90]{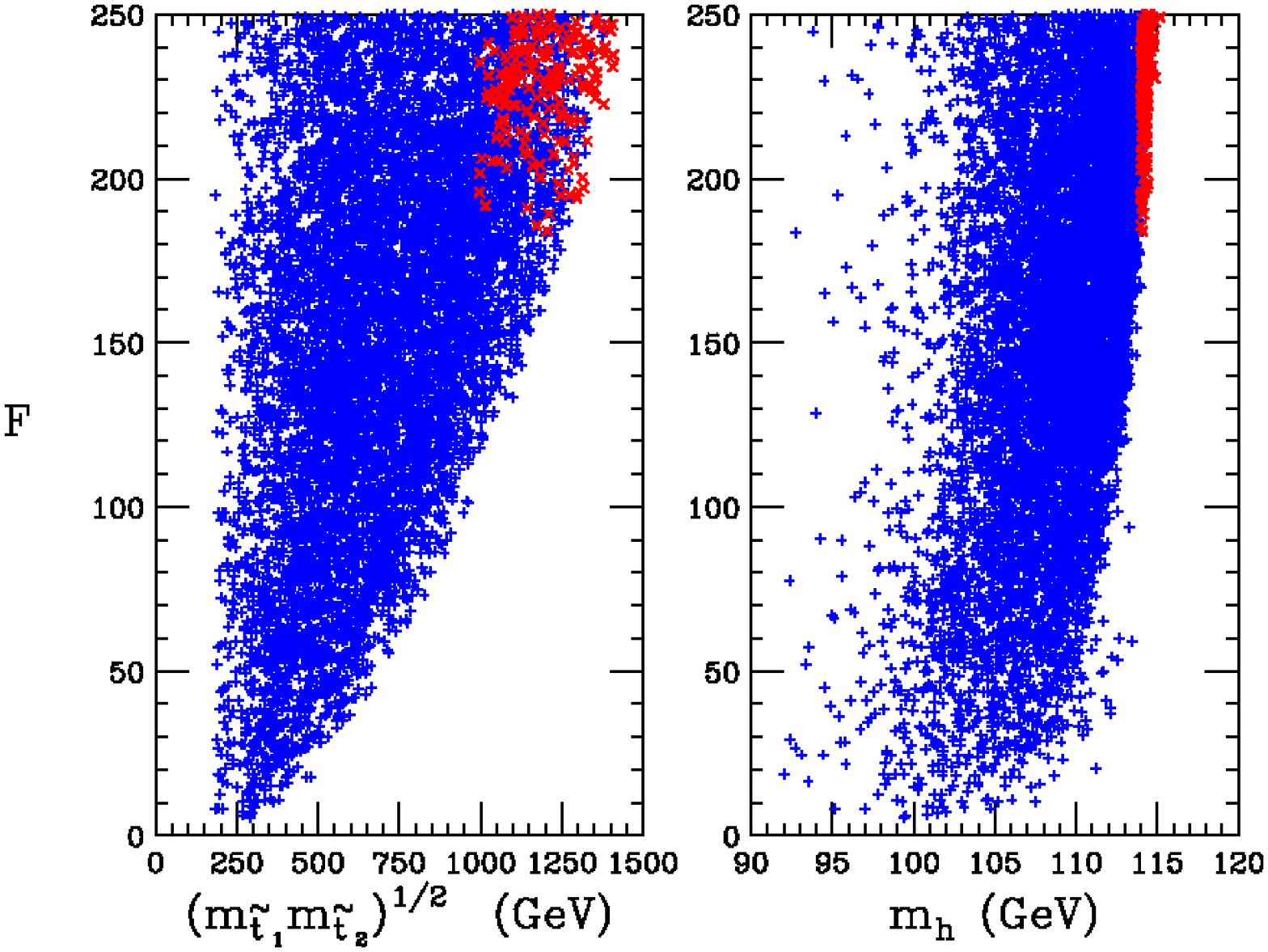}}
\vspace*{-.1in}
\caption{Left: the fine-tuning measure $F$ in the MSSM is plotted vs. $\mstopmean$, without regard to LEP
constraints on $\mh$. The $+$ points have $\mh<114\gev$ and are
excluded by LEP limits.  The $\times$ points
have $\mh>114\gev$ and are experimentally allowed. Right: $F$
is plotted vs. $\mh$ for all scanned points.\vspace*{-.2in}}
\label{mssmnoexpconstraints}
\end{figure}

Results will be presented for $\tanb(\mz)=10$,  
$M_{1,2,3}(\mz)=100,200,300\gev$.
We scan randomly over $|A_t(\mz)|\leq 500\gev$  and 
3rd generation squark and slepton
soft masses-squared above $(200\gev)^2$, 
as well as over $|\mu(\mz)|\geq 100\gev$,
${\rm sign}(\mu)=\pm $ and over $\mha>120\gev$
(for which LEP, MSSM constraints require $\mh\gsim 114\gev$~\cite{lephiggs}).
On the left side of Fig.~\ref{mssmnoexpconstraints},
we plot $F$ as a function of the mean stop mass $\mstopmean$,
which enters into the computation (we use HDECAY~\cite{Djouadi:1997yw}
with $m_t^{\rm pole}=175\gev$) 
of the radiative
correction to the SM-like light Higgs mass $\mh$.
Points plotted as $+$'s ($\times$'s) have
$\mh<114\gev$ ($\mh\geq 114\gev$) and are  excluded 
(allowed) by LEP data.
Very modest values of $F$ (of order $F\sim 5$) are 
possible for $\mh<114\gev$ but
the smallest $F$ value found for $\mh\geq114\gev$
is of order $F\sim 185$~\footnote{Lower $F\gsim 60$ can be
obtained using substantially larger $|A_t|$, which maximizes
$\mh$.}.
The very rapid increase of the smallest achievable 
$F$ with $\mh$ is illustrated
in the right plot of Fig.~\ref{mssmnoexpconstraints}.
This is the essence of
the current fine-tuning problem  for the CP-conserving
MSSM.
Also, to achieve $\mh>114\gev$, $\mstopmean\gsim 1\tev$
is required, an indicator of the little hierarchy problem.

We now contrast this to the NMSSM situation.  
One combination
of the three potential minimization equations yields the 
usual MSSM-like expression
for $\mz^2$ in terms of $\mu^2$, $\tanb$, $\mhusq$
and $\mhdsq$, with $\mu$ replaced by $\mueff$.  However,
a second combination gives an expression for $\mueff$ in terms
of $\mz^2$ and other Higgs potential parameters.
Eliminating $\mueff$, we arrive at an equation of the form $ \mz^4 + 2B \mz^2 +
C = 0$, with solution $\mz^2=-B\pm\sqrt{B^2-C}$, where $B$ and $C$ 
are given in terms of the soft susy breaking parameters, $\lam$,
$\kappa$ and $\tanb$.  Only one
of the solutions to the quadratic equation applies for any given set
of parameter choices. Small fine-tuning is typically achieved when
$C\ll B^2$ and derivatives of $\mz^2$ with respect to a GUT scale
parameter tend to cancel between the $-B$ and $+\sqrt{B^2-C}$
($-\sqrt{B^2-C}$) for $B>0$ (for $B<0$).

To explore fine-tuning, we proceed analogously to the manner
described for the MSSM. At scale $\mz$, we fix $\tanb$ and
scan over values of $\lam\leq 0.5$ ($\lam\lsim 0.7$ is required for perturbativity up to
the GUT scale), $|\kappa|\leq 0.3$, ${\rm sign}(\kappa)=\pm$ and 
$100\gev\leq |\mueff|\leq 1.5\tev$, ${\rm sign}(\mueff)=\pm$.
We also choose $\mz$-scale values for the soft-SUSY-breaking parameters 
$ A_\lam$, $A_\kappa$, $ A_t=A_b$,
$M_1$, $M_2$, $M_3$, $\mqsq$, $\musq$,
$\mdsq$, $\mlsq$, and   $\mesq$, all of which enter into the
evolution equations. We process each such choice through
NMHDECAY (using $m_t^{\rm pole}=175\gev$) 
to check that the scenario satisfies all theoretical and
available experimental constraints (including
$\mstopone\geq 100\gev$).  For accepted cases,
we then evolve to determine the GUT-scale
values of all the above parameters.  The fine-tuning derivative
for each parameter is determined by shifting the GUT-scale
value for that parameter by a small amount, evolving all
parameters back down to $\mz$, redetermining the potential
minimum (which gives new values $\hu^\prime$ and $\hd^\prime$) and 
finally computing a new value for $\mz^2$ using $\mz^{\prime\,2}=\anti
g^{\,2}(
\hu^{\prime\,2}+\hd^{\prime\,\,2})$.

\begin{figure}[h]
\centerline{\includegraphics[width=2.4in,angle=90]{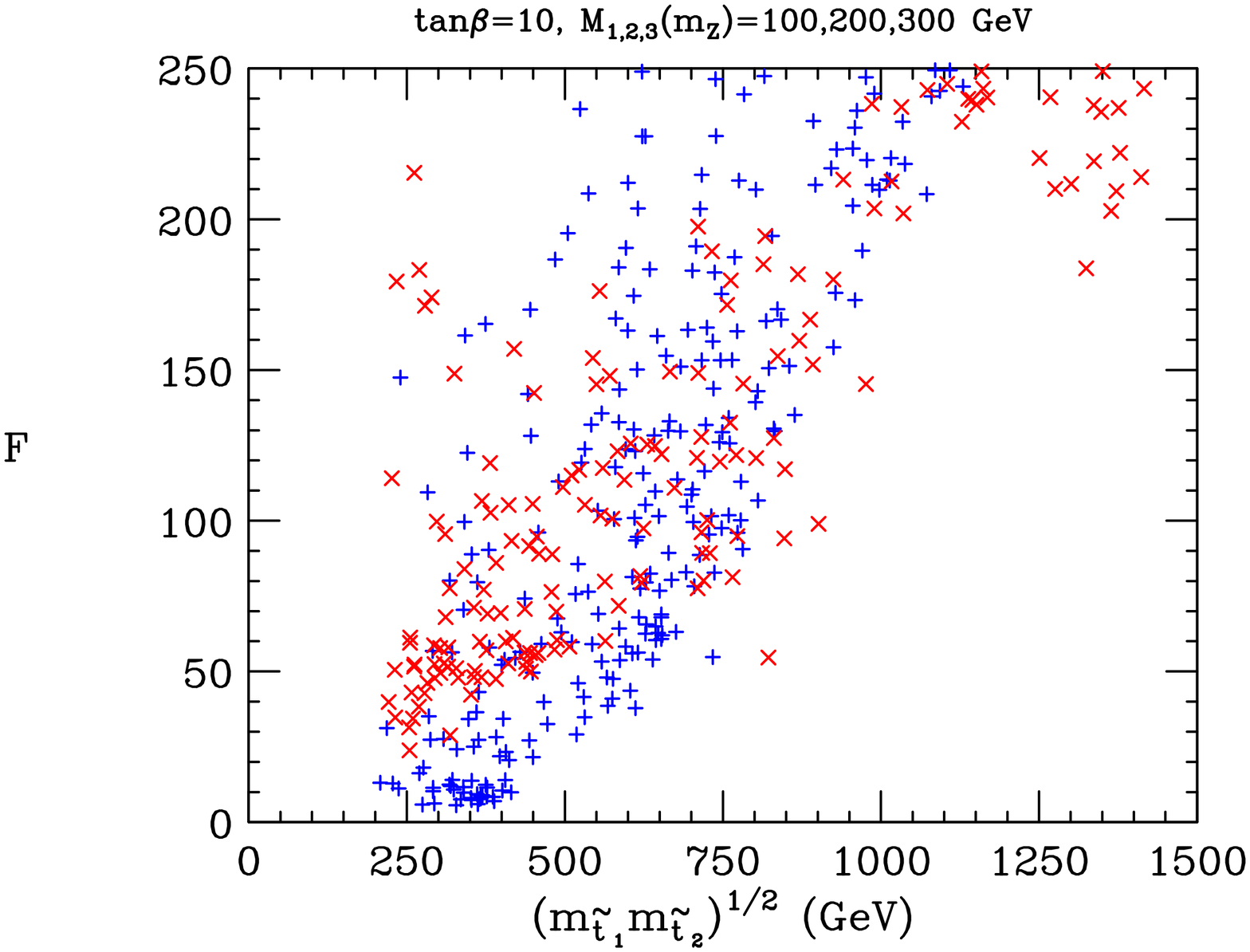}}
\vspace*{-.11in}
\caption{For the NMSSM, we plot the fine-tuning measure $F$
  vs. $\mstopmean$ for NMHDECAY-accepted scenarios
with $\tanb=10$ and $M_{1,2,3}(\mz)=100,200,300\gev$. 
Points marked by '$+$' ('$\times$') escape
LEP exclusion primarily due to dominance of $\hi\to\ai\ai$ decays
(due to $\mhi>114\gev$).\vspace*{-.0in}}
\label{nmssm}
\end{figure}

\begin{figure}[h]
\vspace*{-.12in}
\centerline{\includegraphics[width=2.4in,angle=90]{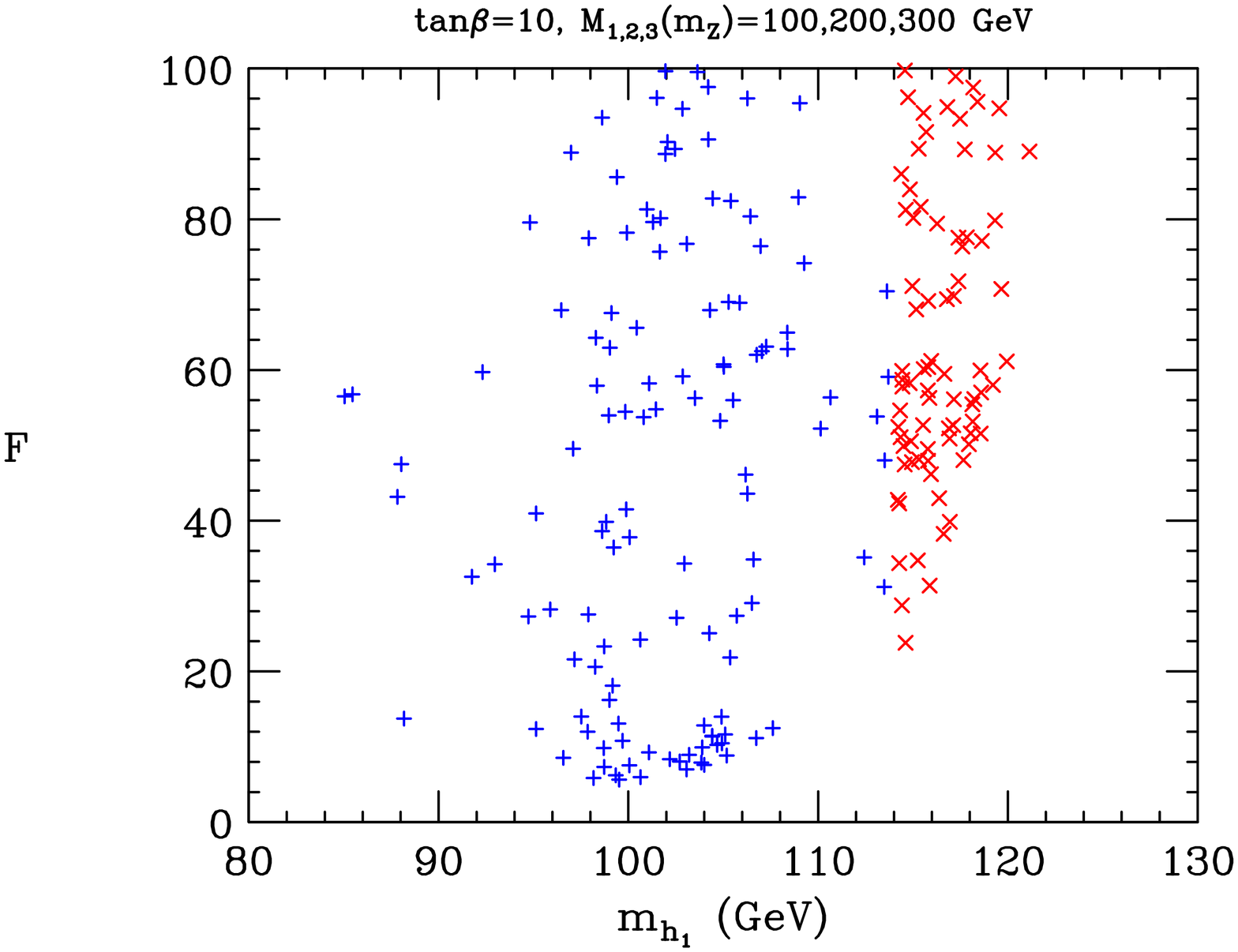}}
\vspace*{-.12in}
\caption{For the NMSSM, we plot the fine-tuning measure $F$ vs. $\mhi$
for NMHDECAY-accepted scenarios
with $\tanb=10$ and $M_{1,2,3}(\mz)=100,200,300\gev$. 
Point labeling as in Fig.~\ref{nmssm}.\vspace*{-.15in}}
\label{nmssmfvsmh}
\end{figure}

Our results for $\tanb=10$ and 
$M_{1,2,3}(\mz)=100,200,300\gev$
and randomly chosen values for the soft-SUSY-breaking
parameters listed earlier 
are displayed in Fig.~\ref{nmssm}. 
We see that $F$ as small as $F\sim 5.5$ can be achieved
for $\mstopmean\sim 250\div 400\gev$.
In the figure, the $+$ points have $\mhi<114\gev$
and escape LEP exclusion
by virtue of the dominance of $\hi\to \ai\ai$
decays; as noted earlier,
 LEP is less sensitive to this channel as compared to the
traditional $\hi\to b\anti b$ decays.  
Points marked by $\times$ have
$\mhi>114\gev$ and will escape LEP exclusion regardless
of the dominant decay mode.  For most of these latter points 
$\hi\to b\anti b$ decays are dominant, even if
somewhat suppressed; 
$\hi\to \ai\ai$ decays dominate for a few. 
For both classes of points,
the $\hi$ has fairly SM-like couplings. We also note that all
points with $F<20$ have $\mhi<114\gev$ and $BR(\hi\to\ai\ai)>0.70$. 
 Finally, in
Fig.~\ref{nmssmfvsmh} we demonstrate the rapid increase of
the minimum $F$ with $\mhi$.  The lowest $F$ values are only
achieved for $\mhi\lsim 105\gev$. However, even for $\mhi\geq 114\gev$,
the lowest $F$ value of $F\sim 24$ is far below that
attainable for $\mh\geq 114\gev$ in the MSSM.

A small value for
$A_\kappa(\mz)$ (typically of order a few $\gev$) 
appears to be essential to achieve small $F$. First, small
$A_{\kappa}$ allows small enough 
$\mai$~\footnote{The small-$A_\kappa$, small-$\mai$
scenario is  well-motivated
by an approximate U(1)$_R$ symmetry~\cite{Dobrescu:2000jt}.} that $\hi\to\ai\ai$
decays are dominant; this makes it possible for
the naturally less fine-tuned
values of $\mhi<114\gev$ to be LEP-allowed. Second, small $F$
is frequently (nearly always) achieved for $\mhi<114\gev$ ($\mhi\geq
114\gev$) via the cancellation mechanism
noted earlier, where $C\ll B^2$, and this mechanism generally works 
mainly for small $A_\kappa$. Indeed,
there are many phenomenologically acceptable parameter
choices with $\mhi>114\gev$ that have large $A_\kappa$, but these all
also have very large $F$.

\begin{figure}[h]
\vspace*{-.07in}
\centerline{\includegraphics[width=2.4in,angle=90]{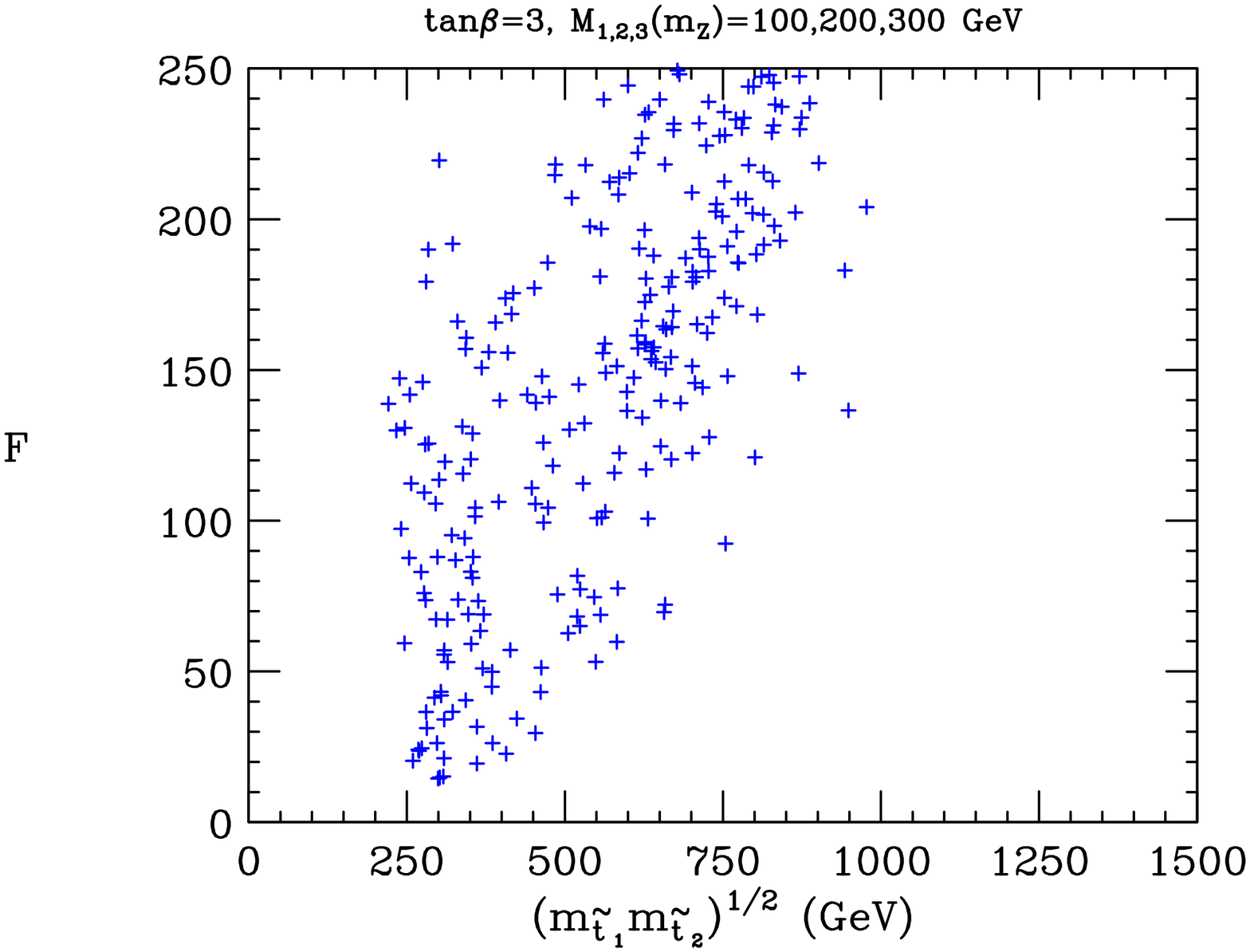}}
\vspace*{-.07in}
\caption{For the NMSSM, we plot the fine-tuning measure $F$ vs. $\mstopmean$ for NMHDECAY-accepted scenarios
with $\tanb=3$ and $M_{1,2,3}(\mz)=100,200,300\gev$. 
Point labeling as in Fig.~\ref{nmssm}.\vspace*{-.1in}}
\label{nmssmtb3}
\end{figure}

For lower $\tanb$ values such as $\tanb= 3$, extremely large
$\mstopmean$ is required for $\mh>114\gev$ in
the MSSM, leading to extremely large $F$.  Results
in the NMSSM for $\tanb=3$ are plotted in Fig.~\ref{nmssmtb3} for
$M_{1,2,3}(\mz)=100,200,300\gev$ and scanning as in the 
$\tanb=10$ case. We see
that $F\sim 15$ is achievable for $\mstopmean\sim
300\gev$.  No points with $\mhi>114\gev$ were found.
All the plotted points escape LEP limits because
of the dominance of the $\hi\to \ai\ai$ decay.
For very large $\tanb$ (\eg\ $\tanb\sim 50$), it is
possible to obtain $\mh>114\gev$ with relatively 
small $\mstopmean$ in the MSSM as well as in the NMSSM.
We have not yet studied fine-tuning at very large $\tanb$
in either model.

In the NMSSM context, the smallest achievable value for $F$ is mainly
sensitive to $M_3(\mz)$.  For example, for $M_3(\mz)\sim 700\gev$
and $\tanb=10$, the smallest $F$ we find is of order
$F\sim 40$. 

We  note that in \cite{bast} 
the mass of the SM-like Higgs $\h$ (where $\h=\hii$
for the parameter choices they focus on)
is increased beyond the LEP limit by
choosing modest $\tanb\sim 2\div 5$ and $\lam$ values 
close to the $0.7$ upper limit consistent with perturbativity
up to the GUT scale. This maximizes the additional NMSSM 
tree-level contribution to $\mh^2$ proportional to $\lam^2$,
thereby allowing $\mh>114\gev$ for somewhat smaller $\mstopmean$ than
in the MSSM.
This, in turn, reduces
the fine-tuning and little hierarchy problems, but not
nearly to the extent achieved by our parameter choices.  
In our plots, the SM-like $\h$ is always the $\hi$. 
The points with 
very small $F$ have low $\mstopmean$, modest 
$\lam$ and $\kappa$, and escape LEP constraints not
because $\mh$ is large but because $h\to aa$ decays
are dominant.

In conclusion, we reemphasize that the NMSSM provides a
rather simple escape from the large fine-tuning and
(little) hierarchy problems characteristic of the CP-conserving
MSSM. However, the relevant NMSSM models  
imply a high probability for $\hi\to\ai\ai$ decays to
be dominant. We speculate that similar results will emerge in
many supersymmetric models where the Higgs sector is more complicated
than that of the MSSM.
Higgs detection in such a decay mode
should be pursued with greatly increased vigor.  Existing work \cite{Ellwanger:2003jt,Ellwanger:2004gz}
which suggests a very marginal LHC signal for $WW\to\hi\to \ai\ai\to
b\anti b \tauptaum$ when $\mai>2m_b$ 
should be either refuted or improved upon.
In addition, the $\ai\ai\to \tauptaum\tauptaum$ channel
that dominates for $2m_\tau<\mai<2m_b$ (an entirely acceptable
and rather frequently occurring mass range in our parameter scans
and not excluded by $\Upsilon$ decays
since the $\ai$ has a large singlet component)
should receive immediate attention. Hopefully, we will not
have to wait for Higgs discovery at an $\epem$ linear collider
via the inclusive $Z\h\to \ell^+\ell^- X$ reconstructed $M_X$
approach (which allows Higgs discovery independent
of the Higgs decay mode) or at a CLIC-based $\gam\gam$ collider
\cite{Gunion:2004si} in the $\gam\gam\to \h\to b\anti b\tauptaum$ or
$\tauptaum\tauptaum$ modes.

\begin{acknowledgments}
This work was supported by the U.S. Department of Energy.
JFG thanks the Aspen Center for Physics where a portion
of this work was performed.
\end{acknowledgments}


\vspace*{-.19in}
\begin{thebibliography}{99}
\vspace*{-.19in}

\bibitem{Arkani-Hamed:2002qy}
N.~Arkani-Hamed, A.~G.~Cohen, E.~Katz and A.~E.~Nelson,
JHEP {\bf 0207}, 034 (2002)
[arXiv:hep-ph/0206021].

\bibitem{Arkani-Hamed:2004fb}
N.~Arkani-Hamed and S.~Dimopoulos,
arXiv:hep-th/0405159.



\bibitem{allg}

H.~P.~Nilles, M.~Srednicki and D.~Wyler,
Phys.\ Lett.\ B {\bf 120} (1983) 346.
%
J.~M.~Frere, D.~R.~T.~Jones and S.~Raby,
Nucl.\ Phys.\ B {\bf 222} (1983) 11.
%
J.~P.~Derendinger and C.~A.~Savoy,
Nucl.\ Phys.\ B {\bf 237} (1984) 307.
%
J.~R.~Ellis, J.~F.~Gunion, H.~E.~Haber, L.~Roszkowski and F.~Zwirner,
Phys.\ Rev.\ D {\bf 39} (1989) 844.
%
M.~Drees,
Int.\ J.\ Mod.\ Phys.\ A {\bf 4} (1989) 3635.
%
U.~Ellwanger, M.~Rausch de Traubenberg and C.~A.~Savoy,
Phys.\ Lett.\ B {\bf 315} (1993) 331 [arXiv:hep-ph/9307322], and
Nucl.\ Phys.\ B {\bf 492} (1997) 21
[arXiv:hep-ph/9611251],
%
S.~F.~King and P.~L.~White,
Phys.\ Rev.\ D {\bf 52} (1995) 4183
[arXiv:hep-ph/9505326].
%
F.~Franke and H.~Fraas,
Int.\ J.\ Mod.\ Phys.\ A {\bf 12} (1997) 479
[arXiv:hep-ph/9512366].


\bibitem{abel1}

S.~A.~Abel, S.~Sarkar and P.~L.~White,
Nucl.\ Phys.\ B {\bf 454} (1995) 663
[arXiv:hep-ph/9506359].

\bibitem{abel2}

S.~A.~Abel,
Nucl.\ Phys.\ B {\bf 480} (1996) 55
[arXiv:hep-ph/9609323],
%
C.~Panagiotakopoulos and K.~Tamvakis,
Phys.\ Lett.\ B {\bf 446} (1999) 224
[arXiv:hep-ph/9809475].

\bibitem{tadp}

H.~P.~Nilles, M.~Srednicki and D.~Wyler,
Phys.\ Lett.\ B {\bf 124} (1983) 337,
%
U.~Ellwanger,
Phys.\ Lett.\ B {\bf 133} (1983) 187,
%
J.~Bagger and E.~Poppitz,
Phys.\ Rev.\ Lett.\  {\bf 71} (1993) 2380
[arXiv:hep-ph/9307317],
%
J.~Bagger, E.~Poppitz and L.~Randall,
Nucl.\ Phys.\ B {\bf 455} (1995) 59
[arXiv:hep-ph/9505244].

\bibitem{radcor1}

U.~Ellwanger,
Phys.\ Lett.\ B {\bf 303} (1993) 271
[arXiv:hep-ph/9302224].

\bibitem{radcor2}

P.~N.~Pandita,
Phys.\ Lett.\ B {\bf 318} (1993) 338,
%
T.~Elliott, S.~F.~King and P.~L.~White,
Phys.\ Rev.\ D {\bf 49} (1994) 2435
[arXiv:hep-ph/9308309],

\bibitem{yeg}

G.~K.~Yeghian,
arXiv:hep-ph/9904488.

\bibitem{higrad2}

U.~Ellwanger and C.~Hugonie,
Eur.\ Phys.\ J.\ C {\bf 25} (2002) 297
[arXiv:hep-ph/9909260].

\bibitem{higsec1}

J.~Kamoshita, Y.~Okada and M.~Tanaka,
Phys.\ Lett.\ B {\bf 328} (1994) 67
[arXiv:hep-ph/9402278],
%
U.~Ellwanger, M.~Rausch de Traubenberg and C.~A.~Savoy,
Z.\ Phys.\ C {\bf 67} (1995) 665
[arXiv:hep-ph/9502206],
%
S.~F.~King and P.~L.~White,
Phys.\ Rev.\ D {\bf 53} (1996) 4049
[arXiv:hep-ph/9508346],
%
S.~W.~Ham, S.~K.~Oh and B.~R.~Kim,
J.\ Phys.\ G {\bf 22} (1996) 1575
[arXiv:hep-ph/9604243],
%
D.~J.~Miller, R.~Nevzorov and P.~M.~Zerwas,
Nucl.\ Phys.\ B {\bf 681} (2004) 3
[arXiv:hep-ph/0304049],
%
G.~Hiller,
arXiv:hep-ph/0404220.

\bibitem{Ellwanger:2004xm}
U.~Ellwanger, J.~F.~Gunion and C.~Hugonie,
arXiv:hep-ph/0406215.


\bibitem{Gunion:1996fb}
J.~F.~Gunion, H.~E.~Haber and T.~Moroi,
arXiv:hep-ph/9610337.

\bibitem{Dobrescu:2000jt}
B.~A.~Dobrescu, G.~Landsberg and K.~T.~Matchev,
Phys.\ Rev.\ D {\bf 63}, 075003 (2001)
[arXiv:hep-ph/0005308].
B.~A.~Dobrescu and K.~T.~Matchev,
JHEP {\bf 0009}, 031 (2000)
[arXiv:hep-ph/0008192].

\bibitem{DELPHI} DELPHI Collaboration,
Eur.\ Phys.\ J.\ C {\bf 38} (2004) 1
[arXiv:hep-ex/0410017].

\bibitem{Opal3} OPAL collaboration, 
Eur.\ Phys.\ J.\ C {\bf 27} (2003) 483
[arXiv:hep-ex/0209068].


\bibitem{Ellwanger:2001iw}
U.~Ellwanger, J.~F.~Gunion and C.~Hugonie,
arXiv:hep-ph/0111179.

\bibitem{Ellwanger:2003jt}
U.~Ellwanger, J.~F.~Gunion, C.~Hugonie and S.~Moretti,
arXiv:hep-ph/0305109.

\bibitem{Ellwanger:2004gz}
U.~Ellwanger, J.~F.~Gunion, C.~Hugonie and S.~Moretti,
arXiv:hep-ph/0401228.


\bibitem{higsec3}

D.~J.~Miller and S.~Moretti,
arXiv:hep-ph/0403137.




\bibitem{bast}

M.~Bastero-Gil, C.~Hugonie, S.~F.~King, D.~P.~Roy and S.~Vempati,
Phys.\ Lett.\ B {\bf 489} (2000) 359
[arXiv:hep-ph/0006198].


\bibitem{King:1995vk}
S.~F.~King and P.~L.~White,
Phys.\ Rev.\ D {\bf 52}, 4183 (1995)
[arXiv:hep-ph/9505326].




\bibitem{finetuning}
G.L.~Kane and S.F.~King, Phys.\ Lett.\ {\bf B451} (1999) 113.

See also: S.~Dimopoulos and G.F.~Giudice, Phys.\ Lett.\ {\bf B357}
(1995) 573; P.H.~Chankowski, J.~Ellis, S.~Pokorski, Phys.\ Lett.\ {\bf
  B423} (1998) 327; P.H.~Chankowski, J.~Ellis, M.~Olechowski,
S.~Pokorski, Nucl.\ Phys.\ {\bf B544} (1999) 39.

\bibitem{lephiggs}
The LEP Working Group for Higgs Boson Searches, LHWG-Note, 2004-01.


\bibitem{Djouadi:1997yw}
  A.~Djouadi, J.~Kalinowski and M.~Spira,
  Comput.\ Phys.\ Commun.\  {\bf 108}, 56 (1998)
  [arXiv:hep-ph/9704448].








\bibitem{Gunion:2004si}
J.~F.~Gunion and M.~Szleper,
arXiv:hep-ph/0409208.



\end{thebibliography}
\end{document}